\documentclass[11pt]{article}
\usepackage{jheppub}

\usepackage[space]{grffile}
\usepackage{orcidlink}
\usepackage{graphicx}
\usepackage[T1]{fontenc} 
\usepackage{amsmath,amssymb,graphicx}
\usepackage{epsf,color}
\usepackage{hyperref}
\usepackage{float}
\usepackage{multirow}
\usepackage{subfigure}
\usepackage[normalem]{ulem}
\usepackage{amssymb}
\usepackage[capitalise]{cleveref}

\definecolor{nicered}{rgb}{0.5,0.,0.}
\definecolor{nicegreen}{rgb}{0.,0.5,0.}
\definecolor{niceblue}{rgb}{0.,0.,0.5}
\hypersetup{colorlinks,citecolor=nicegreen,linkcolor=nicered,urlcolor=niceblue}
\numberwithin{equation}{section}
\newcommand{\beq}{\begin{equation}}
\newcommand{\eeq}{\end{equation}}
\newcommand{\bea}{\begin{eqnarray}}
\newcommand{\eea}{\end{eqnarray}}
\newcommand{\bear}{\begin{eqnarray}}
\newcommand{\eear}{\end{eqnarray}}

\newcommand{\ba}{\begin{array}}
\newcommand{\ea}{\end{array}}

\crefname{section}{Sec.}{Secs.}
\crefname{chapter}{Chapter}{Chapters}
\crefname{figure}{Fig.}{Figs.}
\crefname{equation}{Eq.}{Eqs.}
\crefname{table}{Table}{Tables}
\crefname{appendix}{Appendix}{Appendices}

\title{Non-Standard Neutrino Interactions at a Muon Collider Neutrino Detector}

\date{\today}

\arxivnumber{2508.00761}

\author[a,b,\orcidlink{0000-0002-3100-6144}]{Felix Kling,}
\affiliation[a]{Deutsches Elektronen-Synchrotron DESY, Notkestr. 85, 22607 Hamburg, Germany}
\affiliation[b]{Department of Physics and Astronomy, University of California, Irvine, CA 92697-4575, USA}

\author[c,\orcidlink{0000-0002-9419-6598}]{Yang Ma,}
\affiliation[c]{Centre for Cosmology, Particle Physics and Phenomenology (CP3), UCLouvain, 1348 Louvain-la-Neuve, Belgium}

\author[a,d,e,\orcidlink{0000-0003-4268-508X}]{Krzysztof M\k{e}ka{\l}a,}
\affiliation[d]{Faculty of Physics, University of Warsaw, Pasteura 5, 02-093 Warsaw, Poland}
\affiliation[e]{Center for Particle Physics Siegen, University of Siegen, Walter-Flex-Str. 3, 57072 Siegen, Germany\looseness=-1}

\author[a,\orcidlink{0000-0003-1866-0157}]{J\"urgen Reuter,}
\author[f,\orcidlink{0000-0002-0592-7425}]{Zahra Tabrizi}
\affiliation[f]{{PITT PACC, Department of Physics and Astronomy, University of Pittsburgh, 3941 O’Hara St., Pittsburgh, PA 15260, USA}}

\emailAdd{felix.kling@desy.de}
\emailAdd{yang.ma@uclouvain.be}
\emailAdd{k.mekala@uw.edu.pl}
\emailAdd{juergen.reuter@desy.de}
\emailAdd{z\_tabrizi@pitt.edu}

\preprint{\\DESY 25-112, IRMP-CP3-25-27, PITT-PACC-2506.}

\abstract{
In addition to their broad physics reach enabled by their high energies and precision, future multi-TeV muon colliders will also be the world's most intense sources of neutrinos. This offers the opportunity to search for new non-standard neutrino interactions, possible by installing a dedicated forward neutrino detector in the straight sections of the collision ring, which is then used to measure reactions initiated by neutrinos from the decaying beam muons. In this paper, we show that these searches can exceed current and upcoming bounds on non-standard neutrino interactions from low-energy precision experiments and the LHC. This is achieved by the large flux of high-energetic neutrinos, the precise knowledge of the neutrino flavor composition on each side of the interaction point and the chirality of the neutrinos. We further discuss the technical requirements of the proposed forward neutrino detector, MuCol$\nu$, to maximally exploit this physics potential.}

\makeatletter
\gdef\@fpheader{}
\makeatother

\begin{document}
\maketitle

\section{Introduction}
\label{sec:intro}

In the Standard Model (SM), neutrinos are massless and lepton flavor is conserved due to an accidental global U(1)$_\ell$ symmetry. However, the observation of neutrino oscillations demonstrates violation of lepton flavor and requires neutrinos to be massive, providing evidence for the existence of physics beyond the SM. Despite this, experimental results from precision experiments and direct searches at colliders up to the TeV scale remain in remarkable agreement with SM predictions and, so far, show no evidence for new particles or interactions. This suggests that the scale of new physics responsible for the generation of neutrino masses and violation of lepton flavor lies above the electroweak (EW) scale. 

In this case, the SM serves as an excellent low-energy effective description of a more complete theory, which is augmented by higher-dimensional operators involving the SM fields that emerge from integrating out the heavy degrees of freedom of this theory. This so-called \textit{Standard Model Effective Field Theory} (SMEFT)~\cite{Buchmuller:1985jz, Grzadkowski:2010es} offers a useful framework for describing deviations from the SM predictions. Constraining or detecting non-zero values of the various Wilson coefficients parameterizing this theory is one of the primary goals of current and future experiments. At energy scales far below the EW scale, one can further simplify the theory by integrating out the heavy SM particles, including the Higgs boson, the EW gauge bosons and the top quark, resulting in the \textit{Weak Effective Field Theory} (WEFT)~\cite{Jenkins:2017jig}. A key advantage of this approach is its simultaneous applicability to long-baseline neutrino experiments, including effects of neutrino oscillations.

To look for signs of new physics and tackle the unresolved questions in particle physics, the scientific community plans to expand its toolset by constructing a next-generation collider to succeed the High-Luminosity Large Hadron Collider (HL-LHC) program. The 2020 Update of the European Strategy for Particle Physics~\cite{EuropeanStrategyGroup:2020pow}, the U.S. P5 Report~\cite{P5:2023wyd}, and the 2025 U.S. National Academies report~\cite{NationalAcademiesofSciencesEngineeringandMedicine:2025bix} all emphasize that future efforts should ultimately aim to reach parton collision energies on the order of 10 TeV.
This ambitious objective could potentially be met by a high-energy hadron collider~\cite{FCC:2018vvp, FCC:2025uan}, a linear electron-positron collider using plasma wakefield acceleration~\cite{LinearColliderVision:2025hlt,Gessner:2025acq}, or a circular muon collider~\cite{Accettura:2023ked, InternationalMuonCollider:2025sys}. A key challenge for the muon collider concept is the inherently short lifetime of muons, leading to significant decay rates. However, if the muon collider proves viable, the intense neutrino flux generated from these decays could become a valuable asset, opening the door to a complementary neutrino-focused research program.

Particle colliders produce large numbers of neutrinos, yet for many years, none were directly detected from these sources. This is primarily due to the extremely weak interaction of neutrinos with matter, making their detection in standard collider experiments very challenging~\cite{Foldenauer:2021gkm}. In 2018, the FASER detector was proposed to search for light, feebly-interacting particles in the far-forward region of the ATLAS interaction point~\cite{Feng:2017uoz, FASER:2018ceo, FASER:2018bac, FASER:2018eoc}. Soon after, it was realized that this experiment is also ideally placed to detect and study neutrinos at the LHC~\cite{FASER:2019dxq, FASER:2020gpr, FASER:2021mtu}. As a result, the FASER$\nu$ detector, consisting of emulsion films interleaved with tungsten plates with a mass of about 1~ton, was added to the experiment design~\cite{FASER:2022hcn}. FASER began operation in 2022, reported the first observation of neutrinos in 2023~\cite{FASER:2023zcr} and performed first measurements of neutrino interactions soon after~\cite{FASER:2024hoe, FASER:2024ref}. These efforts have demonstrated the feasibility of neutrino physics measurements at colliders and established the new field of collider neutrino research~\cite{Worcester:2023njy}. 

During Run~3 of the LHC, the FASER$\nu$ experiment is expected to detect approximately 1600 electron neutrinos, 8500 muon neutrinos, and 30 tau neutrinos~\cite{Kling:2021gos, FASER:2024ykc}. Additional measurements are also performed by the SND@LHC experiment~\cite{SNDLHC:2022ihg}. Both FASER and SND@LHC plan to continue operations during the HL-LHC era with upgraded detectors~\cite{FASER:2025myb, Abbaneo:2926288}. Additional neutrino detectors with significantly increased target mass have also been proposed in the context of the Forward Physics Facility~\cite{Anchordoqui:2021ghd, Feng:2022inv, Adhikary:2024nlv, FPF:2025rsc} as well as on the surface exit points of the LHC neutrino beam~\cite{Ariga:2025jgv, Kamp:2025phs}. Detailed studies have identified a broad physics program of such collider neutrino detectors, including measurements of neutrino and muon deep inelastic scattering (DIS) to constrain proton structure at high-$x$~\cite{Cruz-Martinez:2023sdv, Francener:2025pnr}, neutrino flux measurements to constrain forward particle production models, parton structure in unexplored regions at low-$x$, gluon saturation, and intrinsic charm~\cite{Maciula:2022lzk, Bhattacharya:2023zei, Kling:2023tgr, Kling:2025lnt, John:2025qlm} as well as to resolve outstanding problems in astroparticle physics~\cite{Anchordoqui:2022fpn, Bai:2022xad}. Furthermore, these collider neutrinos could be used for searches of neutrino-philic new physics~\cite{Kling:2020iar, Ismail:2020yqc, Falkowski:2021bkq, Kelly:2021mcd, MammenAbraham:2023psg} as well as dark sectors~\cite{Batell:2021blf, Batell:2021aja, Batell:2021snh, Kling:2022ykt}. 

Given the successes of FASER and the broad potential of the collider physics program~\cite{Ariga:2025qup, Kling:2025zon}, it is not surprising that similar detection concepts are being considered for future high-energy colliders. A neutrino experiment at the FCC-hh would benefit from a substantially larger flux, providing the additional opportunity to measure neutrinos from heavy ion collisions or to probe neutrino scattering in polarized target~\cite{MammenAbraham:2024gun}. An even larger neutrino flux is expected at the proposed future multi-TeV muon colliders, which would provide a neutrino beam source with a well-defined composition and energy spectrum through muon decays. As pointed out early on, these high-energy neutrinos could provide insight not only into neutrino cross sections but also allow for the determination of CKM matrix elements, refinement of parton distribution functions (PDFs), the potential discovery of new physics~\cite{King:1997dx, InternationalMuonCollider:2024jyv, Adhikary:2024tvl}, and do precision measurement of the electroweak observables~\cite{deGouvea:2025zfq}.

In this paper, we explore the sensitivity of a future muon collider-based neutrino experiment, dubbed MuCol$\nu$, to potential new physics, focusing on modifications to flavor transition probabilities as described by the WEFT formalism. Many of these physics searches using forward neutrinos complement BSM searches using the main multi-purpose muon collider detectors, e.g. searches for heavy neutral leptons~\cite{Mekala:2023diu,Li:2023tbx, Kwok:2023dck,Mekala:2023kzo,Frigerio:2024jlh,Vignaroli:2025pwn} or $Z'$ bosons~\cite{Korshynska:2024suh}. The structure of the paper is as follows: \cref{sec:detector} outlines a proposed experimental setup at the muon collider for neutrino detection.  \cref{sec:framework} provides an overview of the WEFT formalism. In~\cref{sec:results}, we present our results and compare them with existing complementary constraints. Finally, the conclusions are summarized in~\cref{sec:summary}.

\section{Muon Collider Neutrino Fluxes and Detection}
\label{sec:detector}

\subsection{Neutrino Fluxes and Event Rates}

The muon collider consists of a sophisticated chain of accelerator elements for muon production, cooling, and acceleration. The last stage in this chain is a circular muon storage ring, which delivers the beams to the collision points. Due to their short lifetime, muons decay while circulating in the collider ring, leading to the production of a large number of high-energy neutrinos. As a result, muon colliders act as intense neutrino sources.

\begin{figure}[t]
  \centering
  \includegraphics[trim={0 0 25cm 2cm},clip, width=0.7\textwidth]{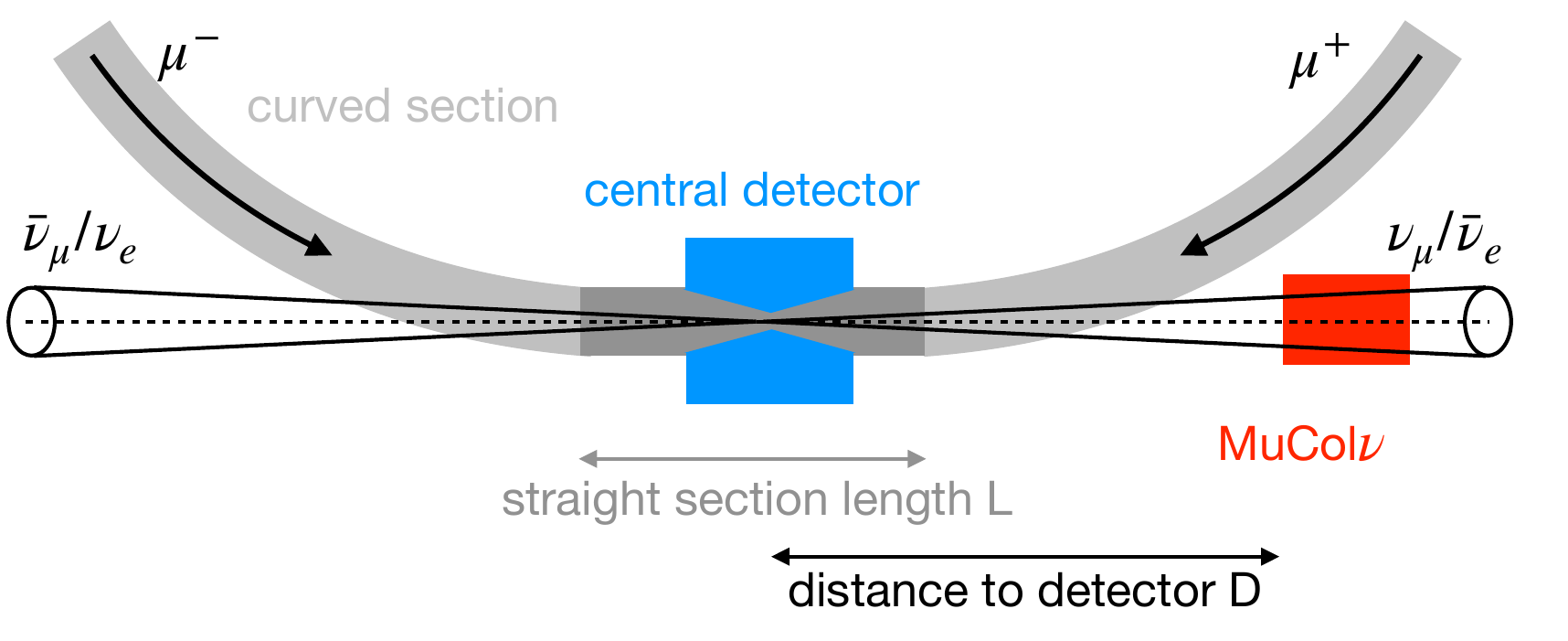}
  \caption{ The muon collider storage ring consists of counter rotating $\mu^+$ and $\mu^-$ beams. Their decays produce intense and strongly collimated beams of neutrinos. The largest flux of neutrinos is produced in the straight sections of the tunnel, close to the main detector, which are assumed to have a length $L$. A dedicated neutrino detector, MuCol$\nu$, is assumed to be placed at a distance $D$ downstream of the interaction point.  }
  \label{fig:location}
\end{figure}

In this study, we consider the neutrino fluxes calculated for a muon collider operating at a center-of-mass energy of 10~TeV, as described in Ref.~\cite{InternationalMuonCollider:2025sys}. The proposed design includes a 10~km long storage ring, into which approximately $10^{13}$ muons are injected every second. While muons decay uniformly throughout the ring, the most intense neutrino beams originate from decays occurring in the straight sections near the interaction points, as illustrated in~\cref{fig:location}. Assuming a straight section of length $L=50$~m, the resulting flux is about $\phi = 5\cdot 10^{10}$ muon neutrinos and electron antineutrinos per second (for the $\mu^-$ beam). Notably, this neutrino flux surpasses the projections for current LHC-based neutrino detectors as well as those proposed for the Forward Physics Facility at the HL-LHC and the FCC-hh by several orders of magnitude, as demonstrated in Figure 2.4.1 of Ref.~\cite{InternationalMuonCollider:2025sys}.

Neutrinos produced from muon decays typically carry transverse momenta comparable to the muon mass and have characteristic energies of several TeV, leading to emission angles of approximately $m_\mu / E_\mu \lesssim 0.1$~mrad. An additional, though still small, angular spread arises from the beam's angular divergence, estimated at about $0.6$~mrad~\cite{InternationalMuonCollider:2025sys}. Owing to the beam's strong collimation, even a relatively compact cylindrical detector with $10$~cm radius positioned at a distance $D=500$~m downstream from the interaction point and aligned with the muon beam collision axis, can achieve a large geometric acceptance of about 10\%. Detectors located farther away, but with proportionally larger transverse area, would be traversed by the same number of neutrinos. 

The cross section for charged-current (CC) neutrino-nucleus interactions rises linearly with the neutrino energy $E_\nu$, and can be approximately written as $\sigma \sim 10^{-35}~\text{cm}^2 \cdot E_\nu/\text{TeV}$. The corresponding anti-neutrino cross section is about a factor of two smaller. The probability of a neutrino interacting with the detector is then given by $P = \frac{\sigma}{A} \ \frac{m}{m_n}$, where $m$ is the detector mass, $m_n$ is the mass of a nucleon, and $A$ is the detector area. Combining the previous results and using a peak energy of $E_\nu \sim 3$~TeV, the expected number of interactions in the detector within a time $t$ is 
\begin{equation}
N \sim 10^5  \times \left(\frac{m}{\text{kg}}\right) \times  \left(\frac{t}{\text{yr}}\right) \times  \left(\frac{L}{50~\text{m}}\right) \times  \left(\frac{500~\text{m}}{D}\right)^2 \ . 
\label{eq:number}
\end{equation}

\begin{figure}
  \centering
  \includegraphics[width=0.7\textwidth]{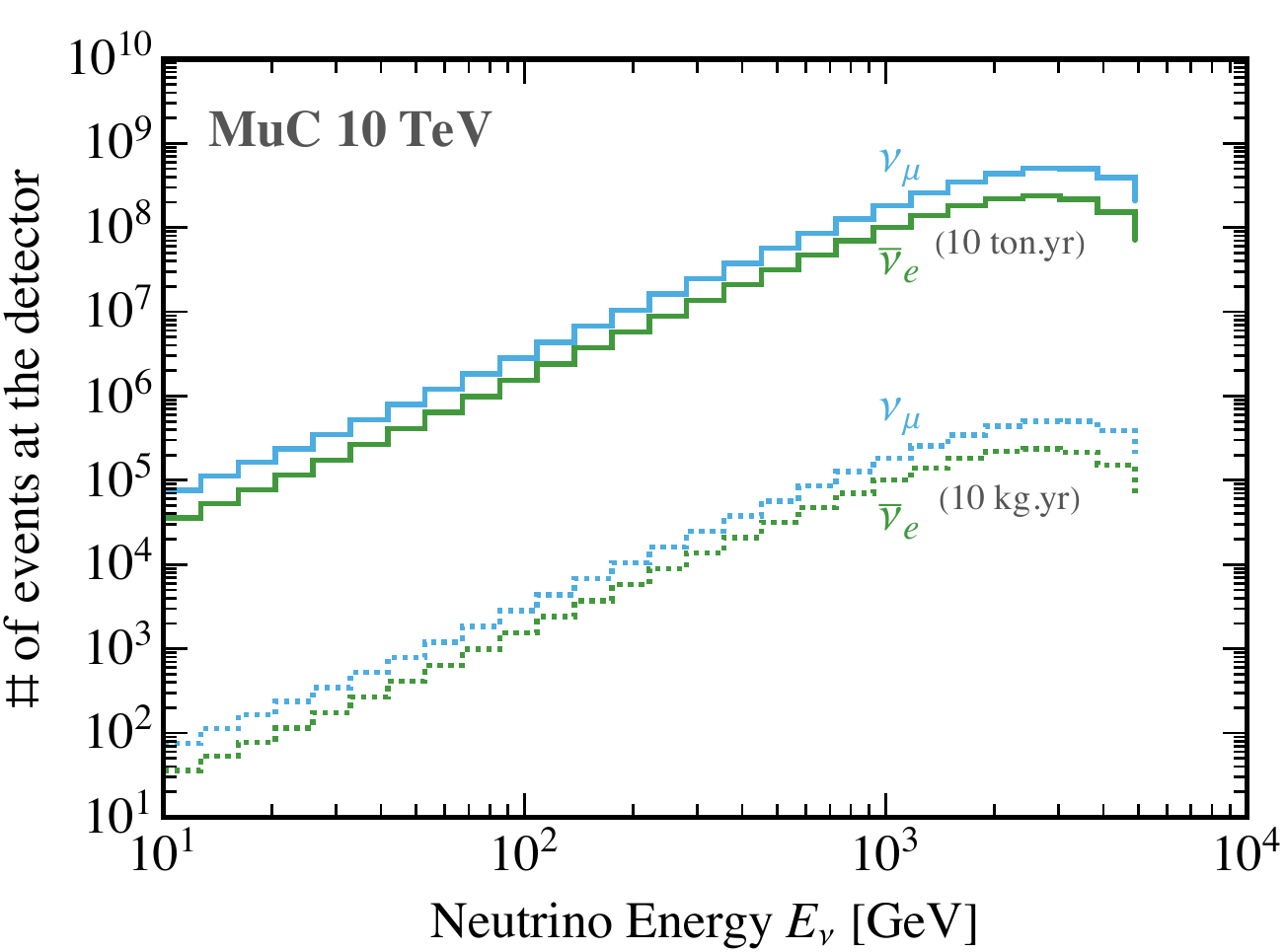}
  \caption{Predicted event spectra at MuCol$\nu$ for the SM at a 10 TeV muon collider, assuming the pessimistic scenario of using a 10~kg$\cdot$yr detector (dotted) and the optimistic case of using the 10 ton$\cdot$yr detector option (solid). 
  }
  \label{fig:spectrum}
\end{figure}

As evident from~\cref{eq:number}, the expected event rate depends strongly on a variety of design parameters: the target mass $m$, the exposure time $t$, the length of the straight section $L$ and distance to the detector $D$. In the above discussion, we have chosen values of $L$ and $D$ consistent with the latest accelerator design, but note that these parameters may still vary by order-one factors in either direction. The value of the target mass, on the other hand, is still completely undetermined. In this study, we follow Ref.~\cite{Adhikary:2024tvl} and consider two representative detector scenarios with exposures of $10~\textrm{kg} \cdot \textrm{year}$ and $10~\textrm{ton} \cdot \textrm{year}$. For example, the latter could correspond to a 1~ton detector operating 10~years or a 10~ton detector operating 1~year. In these scenarios, the expected neutrino interaction rates are approximately 0.1~Hz and 0.1~kHz, corresponding to about $10^6$ and $10^9$ total interactions for the $10~\textrm{kg}\cdot\textrm{year}$ and $10~\textrm{ton}\cdot\textrm{year}$ setups, respectively. The expected number of events at the detector as function of the neutrino energy are shown in~\cref{fig:spectrum} and were obtained by rescaling the fluxes presented in Ref.~\cite{InternationalMuonCollider:2025sys}. We note that the considered spread in target mass also encompasses reasonable variations of the other design parameters, especially $L$ and $D$. 

\subsection{Detector Capabilities}

\begin{figure}[t]
  \centering
  \includegraphics[trim={30cm 3.5cm 0 0},clip, width=0.7\textwidth]{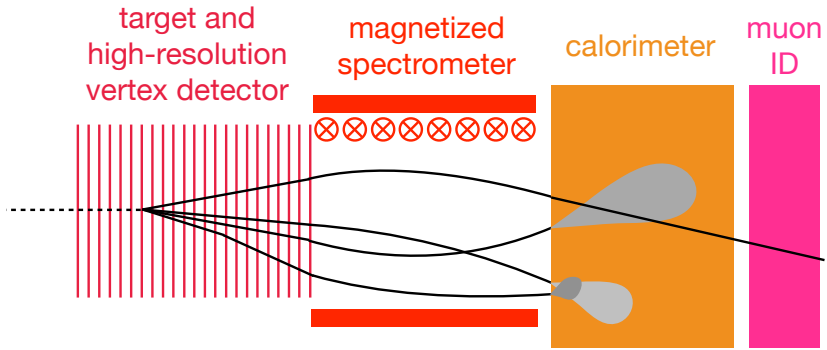}
  \caption{The assumed conceptual design of the MuCol$\nu$ detector follows Ref.~\cite{King:1997dx}. The target consists of a stack of high-resolution tracking planes. These also act as a vertex detector, with the ability to resolve the primary and potential secondary vertices, thereby allowing the identification of short-lived charm hadrons and tau leptons. This is followed by a magnetized spectrometer capable of reliably measuring charge and momentum of multi-TeV energy tracks. Placed at the end is a calorimeter and muon identification system.}
  \label{fig:design}
\end{figure}

There is currently no dedicated design for a forward neutrino detector at a muon collider. However, a potential detector concept for muon storage rings and colliders was previously proposed in Ref.~\cite{King:1997dx}. It consists of a 1~m long cylindrical vertex detector with a 10~cm radius, composed of 750 silicon tracking planes. This is followed by a magnetized spectrometer and a calorimeter, as illustrated in~\cref{fig:design}. This design is intended to provide excellent momentum and energy reconstruction, as well as charged particle identification capabilities, including the tagging of charm and beauty hadrons and $\tau$ leptons. 

In this concept, the tracking planes also serve as the target material, providing a total target mass of about 10~kg. However, incorporating denser materials between the active layers could substantially increase the target mass and, consequently, the expected neutrino interaction rate. Hence, it is feasible to envision more massive detectors with ton-scale targets, similar to the current FASER$\nu$ and SND@LHC detectors at the LHC, or even several hundred tons, akin to the earlier CHARM~\cite{CHARM:1982ods} and NuTeV~\cite{NuTeV:2005wsg} experiments. For simplicity, we assume iron as the target material in our analysis. However, the results primarily depend on the detector mass and are only weakly sensitive to the choice of target nucleus.

The flavor and charge identification abilities play a crucial role for the measurement of flavor-transition inducing WEFT parameters. A neutrino detector placed on one side of the interaction point (IP) at the collider will capture a pure samples of muon neutrinos $\nu_\mu$ and electron antineutrinos $\bar{\nu}_e$, while the corresponding antiparticles, $\bar{\nu}_\mu$ and $\nu_e$, will be produced in the opposite direction from the IP\footnote{In this work, we assume that the detector is positioned along the direction of the muon beam. However, the possibility of deploying a twin detector on the opposite side, enabling the study of interactions involving muon antineutrinos and electron neutrinos, should be considered. Such measurements could help in constraining systematic uncertainties, for example through the sensitivity to a different combinations of underlying quark PDFs, and potentially also provide sensitivity to CP-violating observables.}. In addition, the neutrino oscillation probabilities are negligible for the baseline and energies of a muon collider neutrino experiment. Hence, this directional separation in combination with the detector's ability to identify the final state's lepton flavor and measure its charge in the magnetized spectrometer allows essentially background free searches for flavor transition signatures. 

Finally, let us remark on potential sources of systematic uncertainties. Since the neutrinos originate from beam muon decays, whose trajectory and decays are understood with high precision, the neutrino energy spectrum can be predicted with great accuracy. In addition, for measurements of flavor off-diagonal operators, the neutrino fluxes can be constrained in a data-driven way by measuring the dominant charge current neutrino interaction rates. We therefore assume flux uncertainties to be negligible in our analysis. Another significant source of uncertainty arises from the modeling of the neutrino–nucleus interaction cross section, including uncertainties in PDFs, quark mass effects, higher-order radiative corrections, as well as nuclear shadowing and anti-shadowing effects. A recent prediction based on the NNSF$\nu$ neutrino structure functions introduced in Ref.~\cite{Candido:2023utz} estimates a cross section uncertainty of approximately 3\% for neutrino energies around 1~TeV. Future measurements at facilities such as the FPF~\cite{Cruz-Martinez:2023sdv}, the EIC~\cite{AbdulKhalek:2019mzd, Khalek:2021ulf}, and ultimately MuCol$\nu$, are expected to significantly reduce this uncertainty, bringing it to sub-percent level. In addition, modern multi-purpose Monte Carlo generators are available to simulate neutrino DIS at next-to-leading order, such as POWHEG~\cite{vanBeekveld:2024ziz, FerrarioRavasio:2024kem, Buonocore:2024pdv} and \textsc{Whizard}~\cite{Kilian:2007gr,Bredt:2022dmm,Dahlen:2025udl}.

\section{Theoretical Framework and Experimental Analysis}
\label{sec:framework}

At energies below the electroweak scale, WEFT (similarly named the Fermi theory of four-fermion operators, or the LEFT) is the relevant effective field theory Lagrangian describing the experimental phenomena. This is the case at MuCol$\nu$, with collision center of mass energies $\sqrt{s} = \sqrt{2 E_\nu m_p} \lesssim 40~\text{GeV}$. In this framework, all the heavy particles, including the EW gauge bosons, the Higgs boson, and the top quark are integrated out. The part of the WEFT Lagrangian that we can probe at a dedicated neutrino detector at a high-energy muon collider includes the CC interaction of neutrinos with charged leptons and quarks, which is described by~\cite{Falkowski:2019xoe}:
\begin{align} \label{eq:WEFT_CC}
    \mathcal{L}_{\rm WEFT} \supset & -\frac{2 V_{jk}}{v^2} \bigg\{
        \Big[\mathbf{1} + \epsilon_L^{jk} \Big]_{\alpha\beta}
            \Big(\bar{u}^j \gamma^\mu P_L d^k\Big)
            \Big(\bar{\ell}_\alpha \gamma_\mu P_L \nu_\beta\Big)
      + \Big[\epsilon_R^{jk}\Big]_{\alpha\beta}
            \Big(\bar{u}^j \gamma^\mu P_R d^k\Big)
            \Big(\bar{\ell}_\alpha \gamma_\mu P_L \nu_\beta\Big) \notag\\
    & + \frac{1}{2} \Big[\epsilon_S^{jk} \Big]_{\alpha\beta}
            \Big(\bar{u}^j d^k\Big) \Big(\bar{\ell}_\alpha P_L \nu_\beta\Big)
      - \frac{1}{2} \Big[\epsilon_P^{jk} \Big]_{\alpha\beta}
            \Big(\bar{u}^j \gamma_5 d^k\Big) \Big(\bar{\ell}_\alpha P_L \nu_\beta\Big) \notag\\
    & + \frac{1}{4} \Big[\epsilon_T^{jk}\Big]_{\alpha\beta}
            \Big(\bar{u}^j \sigma^{\mu \nu} P_L d^k\Big)
            \Big(\bar{\ell}_\alpha \sigma_{\mu \nu} P_L \nu_\beta\Big)
      + \text{h.c.} \bigg\}\,.
\end{align}
Here, we have assumed the the EW symmetry is explicitly broken, and $v \equiv (\sqrt{2} G_F)^{-1/2} \approx 246$~GeV is the vacuum expectation value of the SM Higgs field (hence, we do not classify operators into electroweak triplet and singlet operators like e.g. in SMEFT). The mass eigenstates of charged leptons, up-type and down-type quarks are denoted by $\ell_\alpha$, $u^j$ and $d^k$ respectively, while the flavor and mass eigensates of neutrinos are as usual related to each other via the Pontecorvo--Maki--Nakagawa–Sakata (PMNS) mixing matrix $U$: $\nu_\alpha = \sum_{i=1}^3 U_{\alpha i} \nu_i$. Greek indices $\alpha,\beta=e,\mu,\tau$ are used to label lepton flavors, whereas Roman indices $j,k=1,2,3$ denote the mass eigenstates. Moreover, $V_{jk}$  are the enteries of the Cabibbo--Kobayashi--Maskawa (CKM) matrix, $P_{L,R} = \frac{1}{2} (1 \mp \gamma^5)$ are the chirality projection operators, and the tensor $\sigma^{\mu\nu}$ is defined as $\tfrac{i}{2} [\gamma^\mu,\gamma^\nu]$. Finally, subscript $X$ identifies different Lorentz structures corresponding to the left-handed ($L$), right-handed ($R$), scalar ($S$), pseudo-scalar ($P$), and tensor ($T$) interactions, with corresponding dimensionless coefficients $[\epsilon_X^{jk}]_{\alpha\beta}$ that encode the new interactions between quarks and leptons.\footnote{Please note that the sub-indices $\alpha$ and $\beta$ in $[\epsilon_X^{jk}]_{\alpha\beta}$ correspond to $\ell_\alpha$ and $\nu_\beta$, respectively. Therefore, $\epsilon_X$ matrices are not symmetric, and the order of the lepton indices matters.}

\begin{figure}
  \centering
  \includegraphics[width=\textwidth]{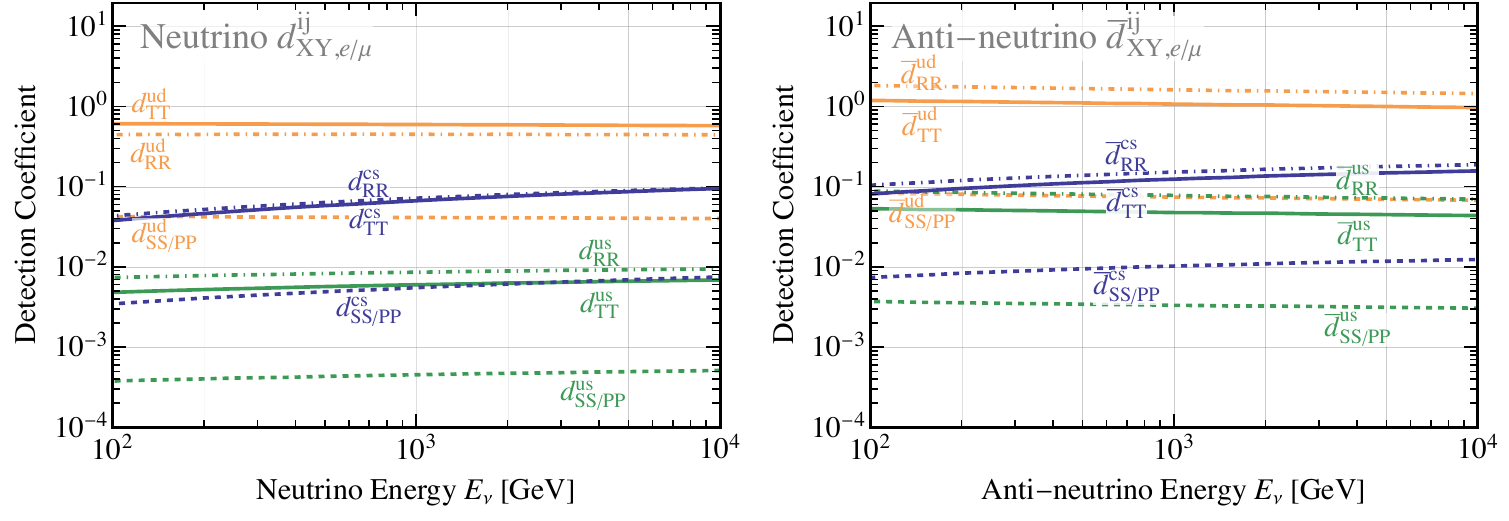}
  \caption{The detection coefficients for the deep inelastic scattering of neutrinos (left) and anti-neutrinos (right) as a function of their energy, based on \cite{Falkowski:2021bkq}. Solid, dot-dashed and dashed curves correspond to tensor couplings, right-handed couplings, and scalar and pseudo-scalar couplings, respectively.}
  \label{fig:dXY}
\end{figure}

The Wilson coefficients in the WEFT Lagrangian could modify the production of neutrinos at the source, the neutrino detection at the target, as well as the neutrino propagation in matter. Concentrating on the CC interactions in~\cref{eq:WEFT_CC}, the corresponding observable would be the number of neutrinos of a given flavor, $N_\beta$, observed at the detector, where the differential event rate reads~\cite{Falkowski:2019kfn, Falkowski:2021bkq}
\begin{align}
  \frac{dN_{\beta}}{dE_\nu} &= n_T\, t\, \sigma_{\beta}^{\rm SM}(E_\nu)
      \sum_{\alpha} \frac{d\phi_{\alpha}^{{\rm SM}}(E_\nu)}{dE_\nu} \,
                      \tilde{P}_{\alpha\beta}(E_\nu)\,.
  \label{eq:rate}
\end{align}
Here, the number of target particles is denoted by $n_T$, the exposure time is shown by $t$ (in units of second), the SM cross section for detecting neutrinos with flavor $\nu_\beta$ at the target is $\sigma_{\beta}^{\rm SM}$ (in units of cm$^2$), and $\phi_{\alpha}^{{\rm SM}}(E_\nu)$ is the flux of neutrinos coming from the source (in units of (cm$^2.$s)$^{-1}$, where we take the value expected in the SM. Finally, $\tilde{P}_{\alpha\beta}$ is the flavor transition probability, which in general accounts for neutrinos oscillations as well as new physics effects that affect neutrino production at the source or neutrino interactions in the detector as derived in Ref.~\cite{Falkowski:2019kfn}\footnote{This probability can be larger than unity as a result of the corrections on the flux and cross section, so in the literature it is often referred to as ``pseudo-probability''.}. Taking into account that neutrino oscillation effects are negligible at MuCol$\nu$ and the considered operators in the WEFT Lagrangian in~\cref{eq:WEFT_CC} do not affect neutrino production in muon decay, we can write the flavor transition probability as~\cite{Falkowski:2019kfn}~\footnote{Please note that the WEFT interactions that describe muon decay include 4-lepton operators, and the corresponding Wilson coefficients are independent from the $\epsilon_X$ in \cref{eq:WEFT_CC}. As the new physics affecting the production and detection of neutrinos at MuCol$\nu$ are independent from each other, we postpone the study of the relevant operators affecting muon decay to a future work. See e.g. Ref.~\cite{Breso-Pla:2023tnz} for the study of EFT affecting muon decay at COHERENT.}
\begin{equation}
  \tilde P_{\alpha\beta}  
  = \delta_{\alpha\beta}
  + \sum_{X,j,k} d_{XL,\beta}^{jk}  [\epsilon_X^{jk}]_{\beta\alpha} \delta_{\beta\alpha}
  + \sum_{X,j,k} d_{XL,\beta}^{jk*} [\epsilon_X^{jk}]^*_{\beta\alpha}
  \delta_{\beta\alpha} 
  + \sum_{X,j,k} d_{XX,\beta}^{jk} \Big|[\epsilon_X^{jk}]_{\beta\alpha} \Big|^2 \, .
  \label{eq:tildeP}
\end{equation}
Here, the ``detection coefficients'' $d_{XX,\beta}^{jk}$ quantify how well a WEFT four-fermion operator with Lorentz structure $X=L,~R,~S,~P,~T$ and quark flavors $jk$ can affect the expected rate of $\nu_\beta$ at the detector. Roughly speaking, the detection coefficients are the ratio between the interaction cross sections and the SM one. The terms $d^{(*)}_{XL}$ describe the interference between a new interaction and the SM amplitude, proportional to one power of $\epsilon$. Instead, the terms with $XX \neq L$ correspond to pure new physics contributions which are proportional to two powers of $\epsilon$. The neutrino detection at MuCol$\nu$ is through the charged-current DIS of neutrinos on a nucleon. The relevant SM cross section, as well as the detection coefficients, are calculated based on Ref.~\cite{Falkowski:2021bkq}. 
As it was shown, in the limit where one can ignore the masses of quarks, the interference with the SM vanishes, and we have  $d_{LX}=0$ for $X \neq L$. This means that for non-SM-like interactions, $X \neq L$, we will be sensitive to the new physics affecting DIS only at the quadratic order with respect to $\epsilon$. This further simplifies~\cref{eq:tildeP} to
\begin{equation}
  \tilde P_{\alpha\beta}  
  \simeq \delta_{\alpha\beta}
  + \sum_{X,j,k} d_{XX,\beta}^{jk}\,\Big|[\epsilon_X^{jk}]_{\beta\alpha} \Big|^2 \, .
  \label{eq:tildeP2}
\end{equation}
The non-zero detection coefficients $d_{XX,\beta}^{jk}$ are shown in~\cref{fig:dXY}, for $\nu_{e,\mu}$ (left panel) and $\bar\nu_{e,\mu}$ (right panel). The detection coefficients for tau (anti) neutrinos are very similar, with slightly different energy dependence. We note that the detection coefficients are usually larger for anti-neutrinos, since they are obtained from the ratio of the new interaction over the SM one, and the SM anti-neutrino cross section is approximately three times smaller. 
Regarding different interactions, we observe that generally the detection coefficients for the right handed and tensor couplings are larger than the scalar and pseudo-scalar ones, where $d_{SS,\beta}^{jk}=d_{PP,\beta}^{jk}$. This is due to small numerical factors that appear in the calculation of these coefficients. On the other hand, regarding the quark flavors, the coefficients are largest for $ud$, thanks to the larger PDFs for the up and down quarks. The $d_{XX,\beta}^{cs}$ coefficients are smaller due to the smaller PDFs for the charm and strange quarks. The $d_{XX,\beta}^{us}$ coefficients are even more suppressed, because of the smaller PDF of strange as well as a suppression by the CKM element $V_{us}^2$. All in all, this means that at MuCol$\nu$ we expect to have the best sensitivity to the $\epsilon_R^{ud}$ and $\epsilon_T^{ud}$ Wilson coefficients. The sensitivity to right handed and tensor coefficients will be a few times (an order of magnitude) worse for the cs (us) quarks. For the scalar and pseudo-scalar coefficients, the sensitivity will be always a few times weaker compared to the $\epsilon_R$ and $\epsilon_T$ for the same quark flavors. 

Having the detection coefficients as well as the relevant fluxes and cross sections, we can rewrite the differential event rate given in~\cref{eq:rate} for MuCol$\nu$. We expect the muon decays to only source muon neutrinos ($\nu_\mu$) and electron anti-neutrinos ($\bar\nu_e$). In the presence of non-standard interactions, we can expect to have an excess of these events with respect to the SM, or they can convert into other neutrino flavors at the detector. For example, the expected rate of $\nu_\mu$ events at the detector will be 
\begin{align}
  \frac{dN_{\mu}}{dE_\nu} &= n_T \sigma_{\mu}^{\rm SM}(E_\nu)
       \frac{d\phi_{\mu}^{{\rm SM}}(E_\nu)}{dE_\nu} \,
                \Big(1+d_{XX,\mu}^{jk} \Big|[\epsilon_X^{jk}]_{\mu \mu} \Big|^2\Big)\,,
  \label{eq:rate2}
\end{align}
which gives us sensitivity to $[\epsilon_X^{jk}]_{\mu\mu}$. On the other hand, $\nu_\mu$ can also convert into $\nu_\beta$ with $\beta=e,\tau$ with the expected rate:
\begin{align}
  \frac{dN_{\beta}}{dE_\nu} &= n_T \sigma_{\beta}^{\rm SM}(E_\nu)
       \frac{d\phi_{\mu}^{{\rm SM}}(E_\nu)}{dE_\nu} \,
                \Big(d_{XX,\beta}^{jk} \Big|[\epsilon_X^{jk}]_{\beta \mu} \Big|^2\Big)\,,
  \label{eq:rate3}
\end{align}
which gives us sensitivity to $[\epsilon_X^{jk}]_{e \mu}$ and $[\epsilon_X^{jk}]_{\tau\mu}$. Similarly, from the detection of $\bar\nu_e$ we will be sensitive to $[\epsilon_X^{jk}]_{ee}$, while in the presence of new physics we can expect to observe $\bar\nu_\mu$ and $\bar\nu_\tau$, giving rise to the $[\epsilon_X^{jk}]_{ \mu e}$ and $[\epsilon_X^{jk}]_{\tau e}$ coefficients. 

In our analysis we consider neutrino energies between $10\,{\rm{GeV}} \leq E_\nu \leq 10^4\,{\rm{GeV}}$, which are sorted into 30 log-spaced bins. We calculate the number of events in each bin of energy while folding the event rates with a Gaussian energy smearing function, considering a width of $0.1 E_\nu$ and a maximum true energy of $E_{\nu}=5\,$TeV (for the 10 TeV muon collider). We assume that the vertex reconstruction efficiency is $100\%$.

We define the following $\chi^2$ function to investigate the sensitivity of MuCol$\nu$ to new physics: 
\begin{align}
  \chi^2(\epsilon_X)
    &= \sum_{\nu,\bar\nu} \sum_{\beta=e,\mu,\tau} \sum_i
       \frac{\big[ N_\beta^i(a, \epsilon_X) - N_\beta^{\text{SM},i} \big]^2}
            {N_\beta^i(a, \epsilon_X)}
     +
     \frac{a^2}{\sigma^2}\,,
  \label{eq:chisq}
\end{align}
where $N_\beta^i(a, \epsilon_X)$ is the expected number of $\nu_\beta$ events as a function of new physics in the $i$-th bin of energy, $N_\beta^{\text{SM},i}$ is the expected SM events, which is non-zero only for $\beta=\mu,\bar{e}$, and zero otherwise. This means that for the non-SM-like events at MuCol$\nu$ the $\chi^2$ simplifies to $N_\beta^i(a, \epsilon_X)$. The parameter $\sigma$ quantifies the systematic uncertainty on the fluxes, with the pull parameter $a$.  For this analysis, we have assumed $1\%$ systematic uncertainty. We have checked that varying this to higher values such as $10\%$ leaves the constraints on off-diagonal $\epsilon$'s unchanged, and only affects the diagonal elements. This is because the sensitivity to the off-diagonal elements is purely statistics-dominated. In order to compute the projected limits, we allow only one of the $[\epsilon_X^{jk}]_{\alpha\beta}$ parameters to be non-zero at a time. Finally, it is important to note that the energy distribution in \cref{eq:chisq} is only relevant for the diagonal elements $[\epsilon_X^{jk}]_{\alpha\alpha}$. In this case, the analysis benefits from the slight dependence of the detection coefficients on the neutrino energy at lower bins, and one becomes more sensitive to these parameters compared to a rate only case. For the off-diagonal coefficients $[\epsilon_X^{jk}]_{\alpha\beta}$ ($\alpha\neq\beta$), where there are no corresponding SM events (see \cref{eq:rate3}), hence the $\chi^2$ becomes a rate-only one and the energy distribution does not help.

Before moving to the results section, it is important to note that in this work we only focus on dim-6 WEFT operators, that correct the number of events as $R = (1 + c \, \epsilon_X^2)R_\text{SM}$, where $R_\text{SM}$ is the SM event rate in the absence of any new physics, while $c$ is a numerical factor (see \cref{eq:rate2}). On the other hand, the contribution from a dim-8 WEFT operator which will be proportional to $\epsilon_8/v^4$ would add the correction $\Delta R = \tilde{c}\, \epsilon_8 (E^2/v^2) R_\text{SM}$ to the total number of events, where $\tilde{c}$ is another numerical factor, and $E$ is the energy scale of the process, in this case the DIS cross section of neutrinos at the detector, that is $E^2\sim x\, E_\nu m_p$ for a given Bjorken parameter $x$. Assuming the dim-6 and dim-8 Wilson coefficients are of the same order, and the numerical factors cancel each other out, it is safe to only keep the dim-6 operators in our calculations as far as the relation  $\epsilon_X \gtrsim E^2/v^2$ is satisfied. We see in the results section that this condition is always true \footnote{See the discussion in Ref.~\cite{Falkowski:2021bkq} for more details}.
\section{Results}
\label{sec:results}

\begin{figure}
  \centering
  \includegraphics[trim={0 2.2cm 0 2.5cm},clip, width=0.9\textwidth]{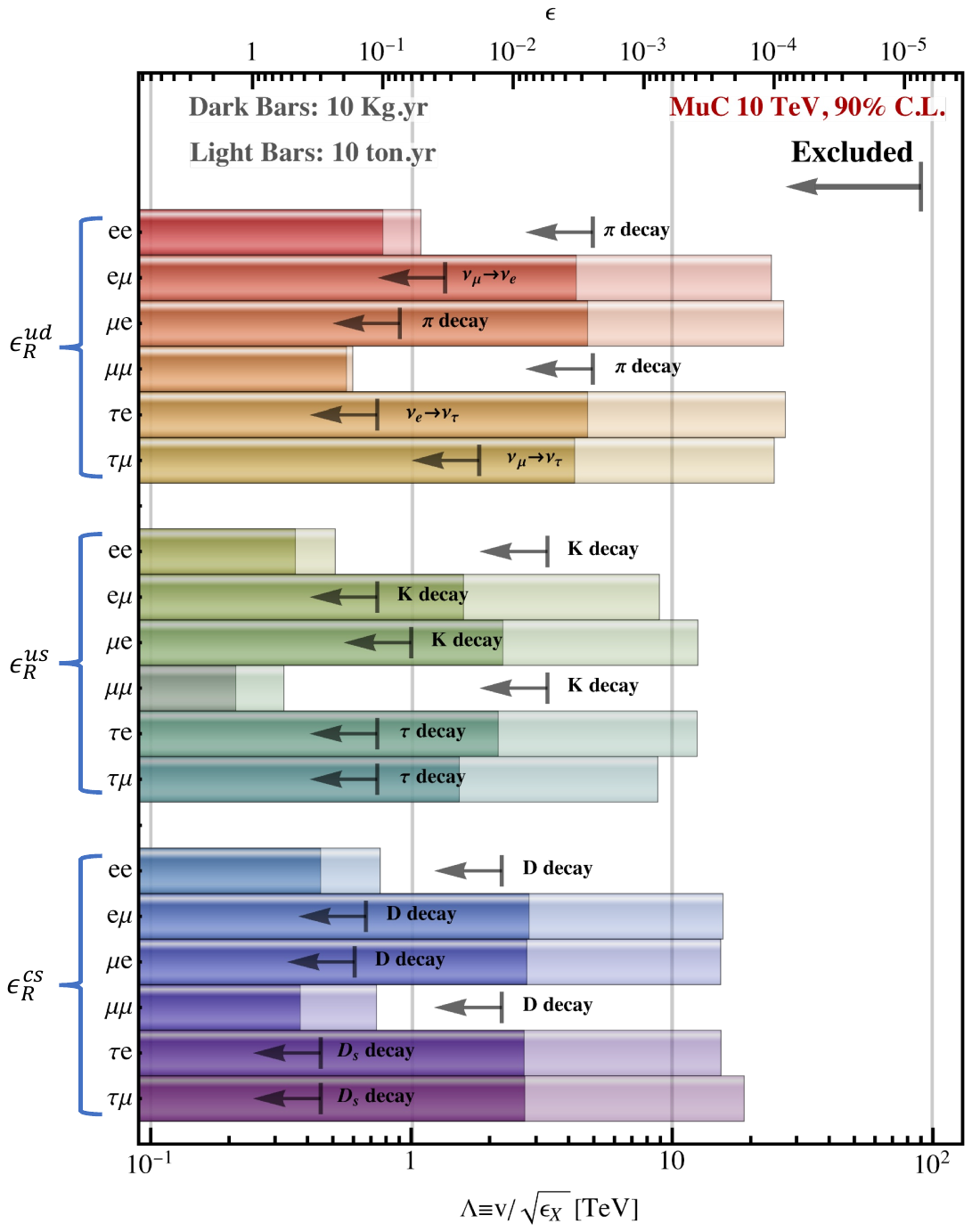}
  \caption{The projected MuCol$\nu$ constraints on the Wilson coefficients of new right-handed interactions in the WEFT framework. Projected limits are reported both in terms of the dimensionless couplings $[\epsilon_X^{jk}]_{\alpha\beta}$ (top axis) and in terms of the effective new physics scale $[\Lambda_X^{jk}]_{\alpha\beta} \equiv v / \sqrt{[\epsilon_X^{jk}]_{\alpha\beta}}$ (bottom axis). Each colored bar indicates the limit on one particular interaction. Darker bars show the projected limits for a 10~kg$\cdot$year detector, while the lighter bars show what a $10~\text{ton}\cdot\text{year}$ neutrino detector can do at a 10 TeV muon collider.}
  \label{fig:constraints-summary-R}
\end{figure}

\begin{figure}
  \centering
  \includegraphics[trim={0 2.2cm 0 2.5cm},clip, width=0.9\textwidth]{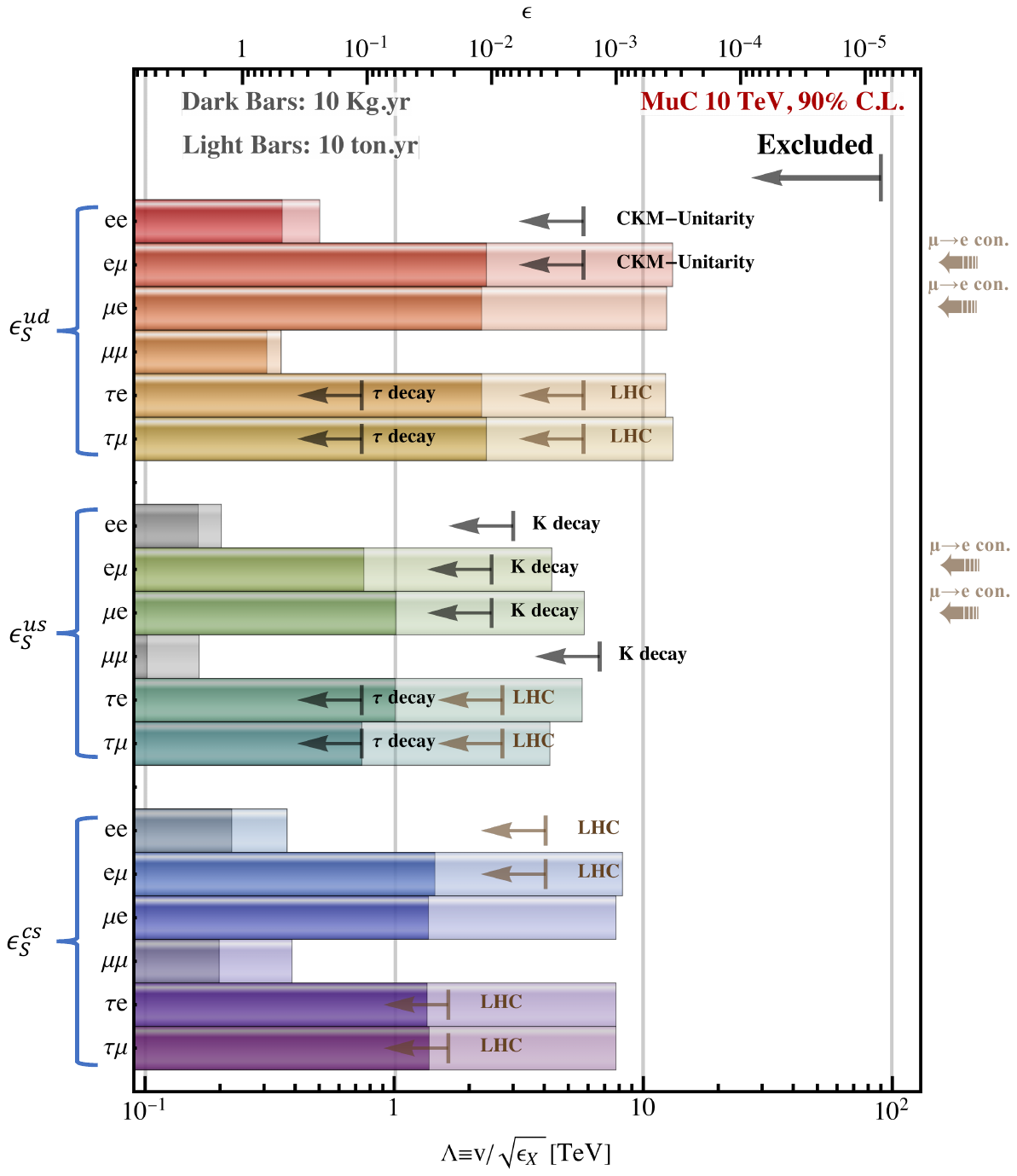}
  \caption{Same as~\cref{fig:constraints-summary-R}, but showing constraints on scalar couplings.}
  \label{fig:constraints-summary-S}
\end{figure}

\begin{figure}
  \centering
  \includegraphics[trim={0 2.2cm 0 2.5cm},clip, width=0.9\textwidth]{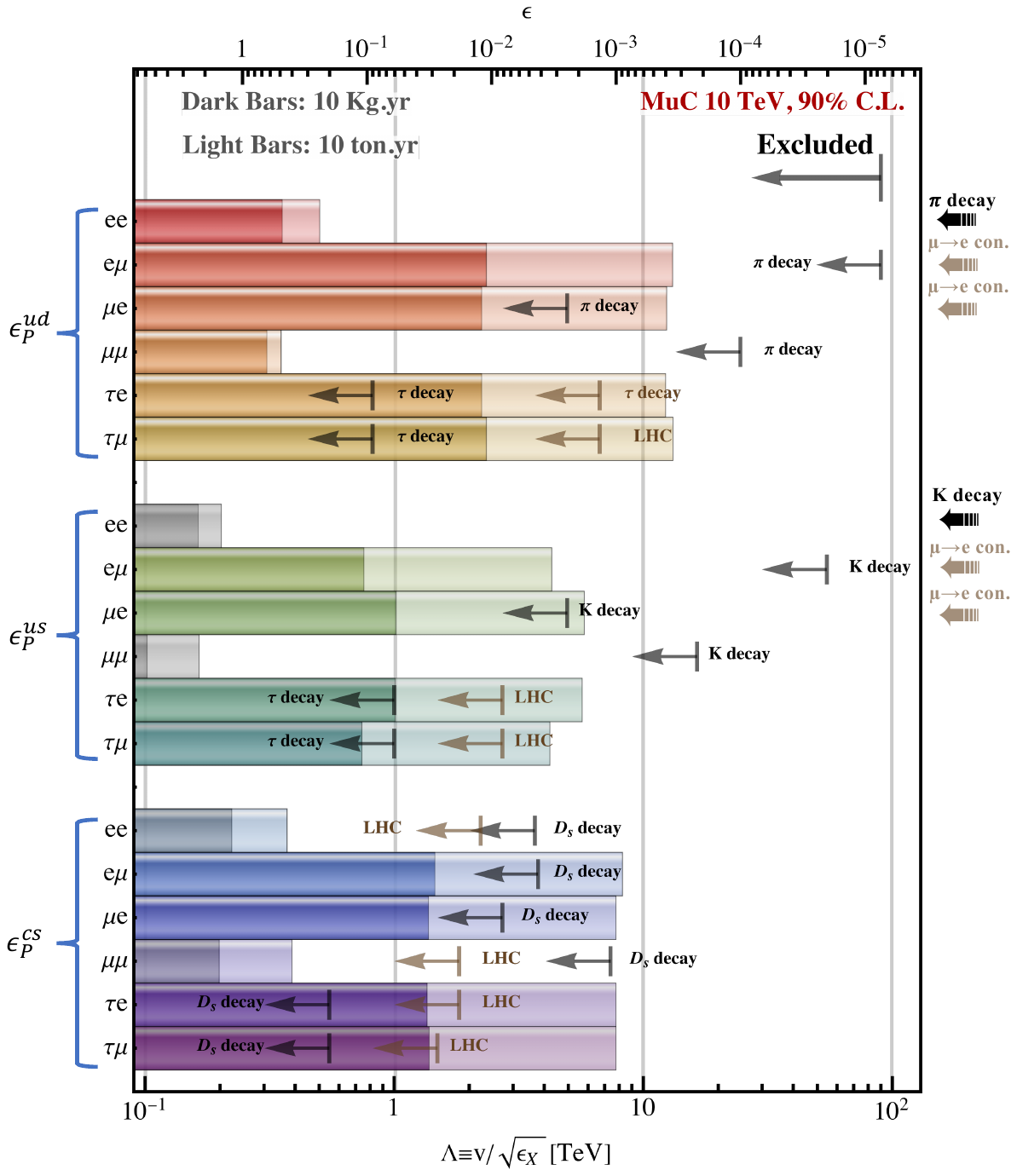}
  \caption{Same as~\cref{fig:constraints-summary-R}, but showing constraints on pseudo-scalar couplings.}
  \label{fig:constraints-summary-P}
\end{figure}

\begin{figure}
  \centering
  \includegraphics[trim={0 2.2cm 0 2.5cm},clip, width=0.9\textwidth]{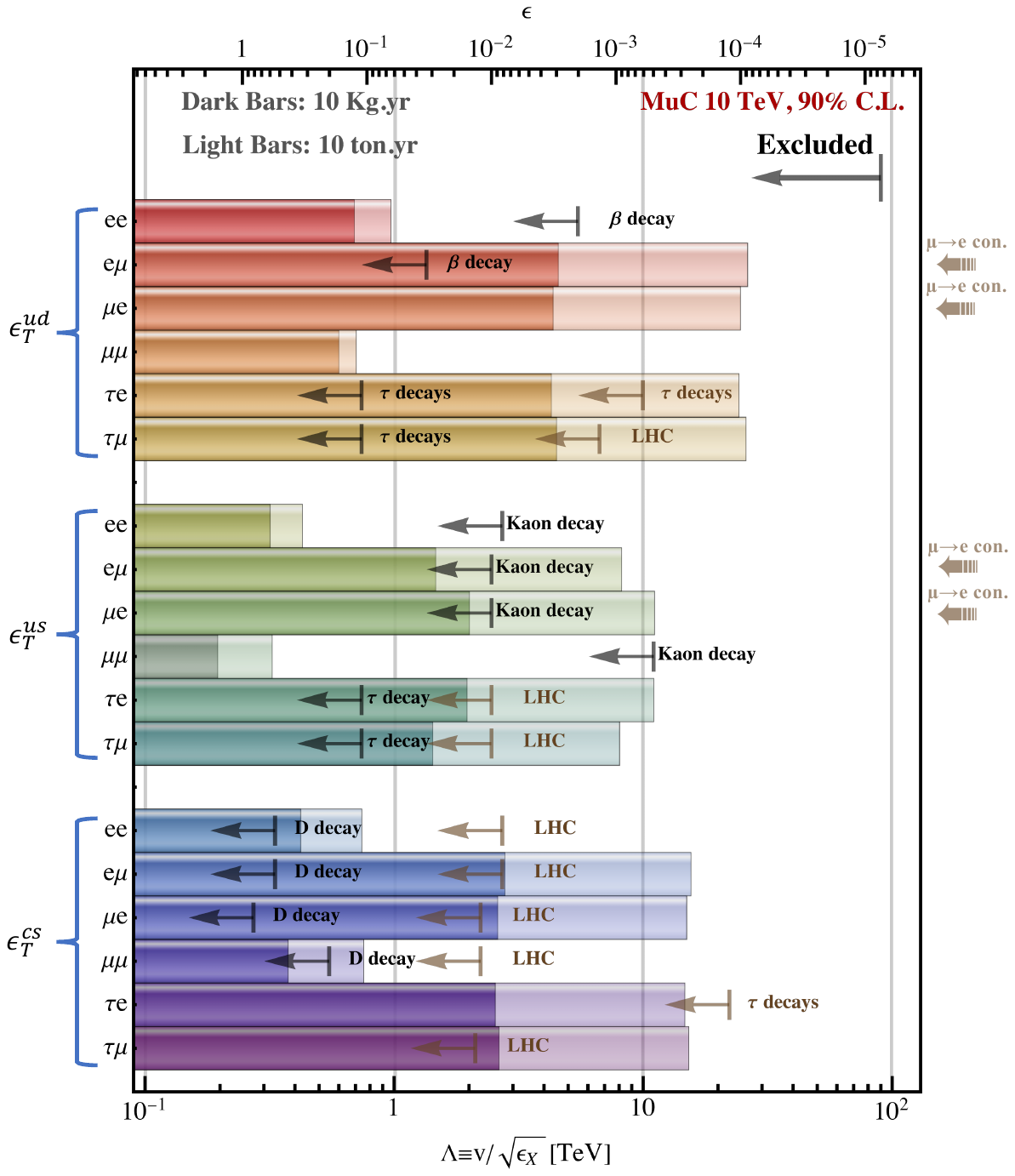}
  \caption{Same as~\cref{fig:constraints-summary-R}, but showing constraints on tensor couplings.}
  \label{fig:constraints-summary-T}
\end{figure}

As discussed in the previous section, the main observable considered is the differential event rate. Observed deviations compared to SM expectations would be a sign of new physics, while their absence allows to constrain the associated WEFT Wilson coefficients. In our analysis we omit the WEFT corrections to left-handed interactions, as the complex interplay of these interactions with the SM processes means that obtaining any meaningful constraint necessitates a re-extraction of the CKM matrix elements from data, accounting for potential new physics contamination. Projected results for the latter are presented for the right-handed operators in~\cref{fig:constraints-summary-R}, the scalar operators in~\cref{fig:constraints-summary-S}, the pseudo-scalar operators in~\cref{fig:constraints-summary-P}, and the tensor operators in~\cref{fig:constraints-summary-T}. In each case, we consider quark flavors $ud$, $us$ and $cs$ and the lepton flavor combinations $ee$, $e\mu$, $\mu e$, $\mu\mu$, $\tau e$, $\tau\mu$, where we remind the readers that a non-vanishing coupling $\epsilon^{jk}_{\alpha\beta}$ induces or modifies the parton level process $\nu_\beta q_j \to \ell_\alpha q_k$. Since the muon collider neutrino beam does not contain $\tau$ neutrinos, no sensitivity for $\epsilon_{\alpha\tau}$ couplings is obtained. Results are presented assuming that only one operator has a non-vanishing value at a time. Projected limits are reported both in terms of the associated couplings $\epsilon$ and the effective new physics scale $\Lambda \equiv v/\sqrt{\epsilon}$.

The best sensitivity is obtained for operators with $e \to \mu$ or $\tau$ as well as $\mu \to e$ or $\tau$ transitions. As mentioned above, the muon collider provides a pure $\nu_\mu$ and $\bar\nu_e$ beam. Hence, no events with primary $\mu^+$, $e^-$ or $\tau^\pm$ are expected to occur in the SM, providing, in principle, a background-free environment to probe lepton flavor-changing processes. This distinguishes a neutrino detector at a muon storage ring from experiments at other neutrino beams, which typically contain a small, yet relevant, fraction of other neutrino flavors.  In addition to operators involving light quarks, MuCol$\nu$ is also sensitive to $\epsilon^{cs}$ which affects the process $\nu s \to \ell c$. In this case, the final-state charm hadron can be tagged via a displaced vertex topology. As explained in the discussion of the detection coefficients, the strange quark PDFs are smaller than those of the up and down quarks, leading to a lower event rate and consequently a reduced sensitivity to $\epsilon^{cs}$ compared to $\epsilon^{ud}$. For the $us$ quarks, there will be an extra suppression due to the CKM element $V_{us}^2$, and hence the reach for $\epsilon^{us}$ is even lower.

The results described above are intended to illustrate the physics potential of a forward neutrino detector at a muon collider and assume the absence of SM background events. However, various processes --- particularly those involving particle misidentification --- could mimic flavor-violating signals. Below, we outline some potential sources of such backgrounds: 
\begin{itemize}
    \item Events with a wrong-sign muon or electron in the final state can arise from displaced pion or kaon decays, such as $\pi^\pm \to \mu \nu$, where the resulting muon is misidentified as prompt. Since the rate of such backgrounds scales approximately linearly with the detector length, using a shorter detector would further suppress them. In principle, precise tracking could help identify these as decay products, although this remains challenging. Prompt wrong-sign leptons may also originate from decays such as $\eta \to \mu^+\mu^-\gamma$, which, however, could be identified by the presence of the second muon. Both sources of wrong-sign muons have been studied in Ref.~\cite{Adhikary:2024tvl} for MuCol$\nu$, which found that the background can be effectively rejected using simple kinematic requirements, while maintaining an $\mathcal{O}(1)$ signal efficiency. At higher target masses and statistics, however, suppressing these backgrounds to negligible levels while maintaining a high signal efficiency becomes more challenging, and improved analysis and background rejection methods will be required.
    
    \item To determine a particle’s momentum and charge in a spectrometer, one relies on the curvature of its track. High-momentum tracks, however, exhibit minimal bending, making charge assignment increasingly difficult and raising the risk of charge mis-assignment. Ensuring reliable charge identification at high energies requires larger and stronger magnets, as well as fine tracking resolution and excellent detector alignment.
    
    \item Tau leptons are typically identified via their displaced decay vertex and their recoil against a hadronic system. However, decays of charm hadrons or pion interactions within the detector material can mimic the secondary vertex of a tau. Reliable tau identification methods have been developed in both collider and neutrino experiments and will need to be adapted and further improved for MuCol$\nu$ \cite{CMS:2022prd, ATLAS:2022aip}.
\end{itemize}

These examples highlight some of the challenges involved in a precision search for flavor-violating signatures. Because the background rates depend sensitively on the specific detector design and performance, we refrain from providing a quantitative estimate at this exploratory stage. The discussed backgrounds are reducible, for example through the methods outlined above, motivating our assumption of an essentially background-free search. Still, accurate particle and charge identification will be critical for this measurement and should be a key design priority for the MuCol$\nu$ detector.

The situation is different for the flavor conserving signatures, where large event rates are expected and the new physics signal would manifest itself only through small distortions of the neutrino energy spectrum. Hence, the sensitivity is significantly reduced, as evident from the $ee$ and $\mu\mu$ entries of~\cref{fig:constraints-summary-R,fig:constraints-summary-S,fig:constraints-summary-P,fig:constraints-summary-T}. We further note that this measurement will be subject to both flux and cross section uncertainties. However, as explained in more detail in~\cref{sec:detector}, flux uncertainties are minimal given our precise knowledge of both the muon trajectories and decays, while cross section uncertainties are currently at the few percent level and are expected to decrease to a sub-percent level by the time of operation of MuCol$\nu$.

In~\cref{fig:constraints-summary-R,fig:constraints-summary-S,fig:constraints-summary-P,fig:constraints-summary-T} we have compared the constraints we expect to get at MuCol$\nu$ with existing constraints from other experiments. The results of low energy experiments, such as precision meson decay experiments, neutrino experiments, and $\beta$-decays can directly be compared with our results, as these experiments have similar energy scales that is relevant for WEFT. We have shown these constraints with the black arrows in~\cref{fig:constraints-summary-R,fig:constraints-summary-S,fig:constraints-summary-P,fig:constraints-summary-T}. On the other hand, the bounds from high energy colliders such as LHC, or charged lepton flavor violation, are obtained within SMEFT. These constraints, shown by brown arrows, are only valid if we assume that WEFT is UV-completed by SMEFT. In summary, the existing constraints relevant for the operators that MuCol$\nu$ is sensitive to are:
\begin{itemize}
    \item The right-handed and pseudo-scalar operators connecting to $ud$ quarks are best constrained by the charged pion decay, as far as the flavor of the observed charged lepton is electron or muon: $[\epsilon_X]_{\alpha\beta}$ for $X=R,P$, $\alpha=e,\mu$ and $\beta=e,\mu,\tau$. This can be achieved by using the ratio $\Gamma(\pi\to e\nu) / \Gamma(\pi\to\mu\nu)$~\cite{Zyla:2020zbs, Cirigliano:2007xi}. Due to the chiral enhancement contribution to the pseudo-scalar interaction, one finds very strong constraints: $\mathcal{O}(10^{-6})$ (for $[\epsilon^{ud}_P]_{e\mu}$) and $\mathcal{O}(10^{-3})$ (for $[\epsilon^{ud}_P]_{\mu e}$)~ \cite{Falkowski:2019xoe}, while the diagonal elements are better constrained, with $[\epsilon^{ud}_P]_{ee}\lesssim10^{-7}$ and $[\epsilon^{ud}_P]_{\mu\mu}\lesssim10^{-4}$~\cite{Gonzalez-Alonso:2016etj}. For the right-handed interaction, the constraints for the diagonal elements are $\mathcal{O}(10^{-3})$~\cite{Falkowski:2021bkq}. 

    \item Neutrino experiments such as the NOMAD experiment~\cite{Astier:2001yj, Astier:2003gs}, looking for flavor transitions, give constraints on $[\epsilon_R^{ud}]_{e\mu,\tau e,\tau\mu}$~\cite{Biggio:2009nt}, which are $\mathcal{O}(10^{-2}-10^{-3})$.

    \item Neutron and nuclear $\beta$ decays give constraints on $[\epsilon_X^{ud}]_{e\beta}$, where they find that the constraints on $[\epsilon_X^{ud}]_{ee}$ is $\mathcal{O}(10^{-3})$ for $X=S,T$~\cite{Falkowski:2020pma}, while  $[\epsilon_{S,T}^{ud}]_{e\mu}$ is constrained to be smaller than $\mathcal{O}(10^{-2})$~\cite{Falkowski:2019xoe}. Using the same $\beta$ decay fit, one can
    use CKM-unitarity and find an upper bound for $[\epsilon_{S}^{ud}]_{ee,e\mu}\leq 2\times 10^{-2}$~\cite{Gonzalez-Alonso:2018omy}.

    \item Operators involving the $us$ quarks, which include $[\epsilon_X^{us}]_{\alpha\beta}$ for $\alpha,\beta=e,\mu$ and with any Lorentz structure, are best constrained by the leptonic kaon decay, by using the ratio $\Gamma(K\to e\nu) / \Gamma(K\to\mu\nu)$~\cite{Falkowski:2021bkq}. The corresponding bounds are $\mathcal{O}(10^{-3}-10^{-6})$ for pseudo-scalar couplings and $\mathcal{O}(10^{-1}-10^{-3})$ for the other interactions~\cite{Falkowski:2021bkq}.

    \item Hadronic $\tau$ decays are sensitive to $[\epsilon_X^{jk}]_{\tau\beta}$ operators, for $\beta=e,\mu,\tau$ and all possible Lorentz structures. 
    For example, for the coefficients with $ud$ quarks the corresponding bounds can be obtained from the $\tau \to \pi \nu$ decays, while for the $us$ couplings one can use the ratio of $\Gamma(\tau \to K \nu)/\Gamma(K \to \mu \nu)$. The constraints on the off-diagonal Wilson coefficients are in the range $10^{-2}-10^{-3}$, as the decay rate is sensitive to them only in quadratic order~\cite{Falkowski:2021bkq}. 

    \item For operators that include $cs$ quarks, constraints can be obtained from leptonic and semi-leptonic $D$ meson decays. The constraints are strongest for the $ee,e\mu,\mu e,\mu\mu$ elements of the pseudo-scalar coupling, and are $\mathcal{O}(10^{-3})$~\cite{Falkowski:2021bkq}, and they are an order of magnitude weaker for the right handed interactions. For the $\tau e$ and $\tau \mu$ elements of $R$ and $P$ the upper bounds are $\mathcal{O}(10^{-1})$~\cite{Falkowski:2021bkq}. Finally, the $ee,e\mu,\mu e,\mu\mu$ elements of the tensor coupling are constrained to be $\mathcal{O}(1)$~\cite{Falkowski:2021bkq}. 
    
    \item If we assume that WEFT is UV completed by SMEFT, we can use LHC constraints for comparison. In this case, one needs to consider the running and matching between the SMEFT and WEFT parameters. Using the Drell-Yan process at LHC one can obtain $\mathcal{O}(10^{-2}-10^{-3})$ bounds on most WEFT operators with $X=S,P,T$~\cite{Falkowski:2021bkq}. 

    \item Finally, the experimental bounds on  $\mu \to e$ conversion on gold nuclei~\cite{Bertl:2006up} place very strong constraints on $[\epsilon_X^{jk}]_{e\mu,\mu e}$ operators for $X=S,P,T$ and for $ud$ and $us$ quarks, which are $\mathcal{O}(10^{-6}-10^{-8})$~\cite{Falkowski:2021bkq}. However, we stress again that these constraints do not hold if WEFT is not UV-completed by SMEFT, e.g. if the new physics exists below the electroweak scale. 

    \end{itemize}

\section{Conclusions}
\label{sec:summary}

The SM is known to be incomplete, and the observation of neutrino oscillations, demonstrating lepton flavor violation, may serve as a gateway to new physics. In the absence of direct experimental evidence for physics beyond the SM, more general theoretical frameworks, such as the Weak Effective Field Theory (WEFT), offer a systematic approach to explore potential deviations from the SM. In this work, we examine how various effective operators could be probed at a hypothetical neutrino experiment at a future muon collider, envisioned as the most intense artificial source of neutrinos ever conceived.

This experiment, called MuCol$\nu$ in this study, is expected to detect neutrinos at kHz rates, enabling an unprecedented level of precision in neutrino scattering measurements. Such a capability would allow for highly accurate constraints on charged-current neutrino-nucleon interactions. As demonstrated in our analysis, the MuCol$\nu$ detector has the potential to deliver world-leading sensitivity to a variety of operators within the WEFT framework, featuring different Lorentz structures, in particular those involving lepton flavor violation. We find that for the right-handed and tensor operators with $ud$ quarks, MuCol$\nu$ can obtain constraints $\mathcal{O}(10^{-4})$, translating to new physics reach at about 30 TeV. For several four-fermion operators, the expected sensitivity at MuCol$\nu$ will be up to two orders of magnitude improved compared to the existing constraints. Most importantly, for various interactions, MuCol$\nu$ will be able to surpass the constraints from kaon decay experiments, which are designed for precision tests.

These considerations strongly support integrating the MuCol$\nu$ as a core component of the muon collider physics program from its beginning. To ensure optimal sensitivity to the discussed WEFT operators, the following detector requirements should be fulfilled:

\begin{itemize}
\item While SM processes only produce $\mu^-$ and $e^+$ in neutrino interactions, the observation of opposite-sign leptons would signify lepton-number-violating processes. Reliable charge identification for leptons at high energies is essential for the considered measurements.

\item In the SM, the appearance of tau neutrinos in this setup is not expected. Their observation would provide a clear signal of new physics. Therefore, the detector should be capable of identifying tau neutrinos with high efficiency.

\item The detector should be capable of efficiently tagging charm hadrons, which can play a key role in probing effective operators and interaction channels involving the charm quarks.

\item Precise control over background sources is crucial. This includes, for example, suppressing the production of opposite-sign muons from in-flight $\pi^+ \to \mu^+\nu$ decays and their misidentification of prompt muons as well as minimizing mis-tagging of scattering particles as tau-like secondary decay vertices.

\item The placement of a twin detector on the opposite side of the interaction point could enhance control over systematic uncertainties and enable complementary measurements, possibly including the probing of CP-violating observables. 
\end{itemize}

While this work focused on physics modifying charged-current interactions affecting neutrino detection, MuCol$\nu$ can also probe neutral-current neutrino interactions. Although more challenging to analyze due to the undetected final-state neutrino, these are known for their sensitivity to non-standard interactions~\cite{Ismail:2020yqc}, neutrino charge radii~\cite{Mathur:2021trm}, or the weak mixing angle~\cite{MammenAbraham:2023psg}. Beyond this, MuCol$\nu$ could serve for a broad range of physics goals, including improved determinations of PDFs, precise measurements of CKM matrix elements, sterile neutrino oscillations, and other dedicated searches for signals of new physics. Insights from this complementary set of studies should be considered in the detector design to ensure maximum scientific reach.\\

\acknowledgments

We thank nature for containing muons, acknowledging their indispensable role in enabling this study. We further thank Dario Buttazzo, Tova Holmes, Joachim Kopp, Lawrence Lee, Patrick Meade, Federico Meloni, Mark Palmer, Daniel Schulte, Sebastian Trojanowski and Andrea Wulzer for useful discussions about muons and how to collide them. 
The work of K.~M. and J.R.~R. has been supported by the Deutsche Forschungsgemeinschaft (DFG, German Research Foundation) under Germany’s Excellence Strategy-EXC 2121 “Quantum Universe”-390833306 and by the National Science Centre (Poland) under OPUS research project no. 2021/43/B/ST2/01778 a.
This work has also been funded by the DFG grant 491245950.
K.~M. additionally acknowledges the support of the Polish National Agency for Academic Exchange (NAWA), Poland.
The work of F.~K. is supported in part by Heising-Simons Foundation Grant 2020-1840 and in part by U.S. National Science Foundation Grant PHY-2210283.
Y.~M. acknowledges the support as a Postdoctoral Fellow of the Fond de la Recherche Scientifique de Belgique (F.R.S.-FNRS), Belgium.
Y.~M. is also partially supported by the IISN convention 4.4517.08, ``Theory of fundamental interactions.''
The work of Z.~T. is supported by Pitt PACC. 
We thank the Galileo Galilei Institute for Theoretical Physics for the hospitality and the INFN for partial support during the completion of this work.

\bibliographystyle{JHEP}
\bibliography{ref.bib}

\end{document}